\documentclass{emulateapj}
\usepackage{graphicx}
\usepackage{amssymb}
\usepackage{amsmath}
\usepackage{natbib}
\usepackage{hyperref}

\shorttitle{Chandra pulsar survey (ChaPS).}
\shortauthors{Kargaltsev et al.}

\newcommand{\edot}{\dot{E}}

\newcommand{\lpsrxc}{$L_{\rm X, crit}^{ \rm psr}$}
\newcommand{\lpwnxc}{$L_{\rm X, crit }^{ \rm pwn}$}
\newcommand{\lpsrx}{$L_{\rm X}^{ \rm psr}$}

\newcommand{\lpsrg}{$L_{\gamma}^{ \rm psr}$}

\newcommand{\etapwnx}{$\eta_{\rm X}^{\rm pwn}$}
\newcommand{\etapsrx}{$\eta_{\rm X}^{\rm psr}$}

\newcommand{\chan}{{\sl CXO} }
\newcommand{\be}{\begin{equation}}
\newcommand{\ee}{\end{equation}}

\begin{document}


\title{Chandra pulsar survey (ChaPS).}

\author{Oleg Kargaltsev\altaffilmark{1}, Martin Durant\altaffilmark{1}, 
 George G.\ Pavlov\altaffilmark{2,3}, and Gordon Garmire\altaffilmark{2}}

\altaffiltext{1}{Department of Astronomy, University of Florida, Gainesville, FL 32611--2055, USA; martin.durant@astro.ufl.edu, oyk100@astro.ufl.edu}
\altaffiltext{2}{Department of Astronomy and Astrophysics, Pennsylvania State University,
University Park, PA 16802, USA; pavlov@astro.psu.edu}
\altaffiltext{3}{St.-Petersburg State Polytechnical University, Polytekhnicheskaya ul.\ 29, 195251, Russia}


\begin{abstract}
Taking  advantage of the high sensitivity of the \chan Advanced CCD Imaging Spectrometer,
 we have conducted a snap-shot survey of 
  pulsars
  previously undetected in X-rays. We detected 12 pulsars  
 and established deep flux limits for  11 pulsars. Using  these new results, we revisit the relationship between the X-ray luminosity, \lpsrx, and spin-down power, $\edot$. We find that 
 the obtained limits further 
 increase
   the extremely large spread in the non-thermal X-ray efficiencies, \etapsrx=\lpsrx/$\edot$, with some of them being now below $10^{-5}$. Such a spread cannot be explained by  poorly known distances or by beaming of pulsar radiation.  We also find  evidence 
  of a break in the dependence of  \lpsrx on $\edot$,  such that pulsars become more X-ray efficient at $\dot{E} \lesssim 10^{34}$--$10^{35}$ erg s$^{-1}$. 
  We examine 
 the relationship between the $\gamma$-ray luminosity, \lpsrg, and $\dot{E}$, which exhibits a smaller scatter compared to that in X-rays. This confirms that the very large  spread in the X-ray efficiencies cannot be explained just by the beaming because the $\gamma$-ray emission is generally expected to be beamed stronger than the X-ray emission.  Intriguingly, there is also an indication of  a break in the \lpsrg$(\edot)$ dependence  at $\edot\sim10^{35}$ erg s$^{-1}$, with lower-$\edot$ pulsars becoming less $\gamma$-ray efficient.
   We also examine the distance-independent \lpsrg/\lpsrx ratio as a function of $\edot$ for a sample of $\gamma$-ray pulsars 
  observed by \chan\/ and  find  that it peaks at $\edot\sim10^{35}$ erg s$^{-1}$, showing that the 
    breaks cannot originate from  poorly measured distances. We discuss the implications of our findings for existing models of magnetospheric emission and venues for further exploration.
\end{abstract}
\maketitle

\section{Introduction}

Along with observations in the radio band, X-rays have been the primary spectral window to study rotation-powered pulsars.
Thanks to their superb sensitivity and angular resolution,  the  latest generation  X-ray telescopes
 have detected emission from $\gtrsim100$ rotation-powered isolated pulsars 
  and  about 60 pulsar-wind nebulae (PWNe; \citealt{2008AIPC..983..171K,2010AIPC.1248...25K}; hereafter KP08 and KP10).  
   The growing sample allows one to look for dependences between the X-ray properties and other pulsar parameters, such as  the pulsar rotational energy loss rate    (spin-down power) $\dot{E}$.  

There have been multiple  attempts to establish the 
relationship between the pulsar X-ray luminosity, \lpsrx, and $\dot{E}$, 
including
 those
  based on  {\sl Einstein} data (\lpsrx$\propto \edot^{1.39}$; \citealt{1988ApJ...332..199S}),  {\sl ROSAT} data (\lpsrx$\sim 10^{-3} \edot$; \citealt{1997A&A...326..682B}),  {\sl ASCA}, {\sl RXTE}, {\sl BeppoSAX}, \chan, and {\sl XMM-Newton}  data (\lpsrx$\propto\edot^{1.34}$; \citealt{2002A&A...387..993P}),   \chan\/ and {\sl XMM-Newton} data  (\lpsrx$\propto\edot^{0.92}$; \citealt{2008ApJ...682.1166L}), and \chan\/ data (KP08).

One of the reasons for such a variety of scaling relations is the different approaches used by different authors. For instance, they used different energy ranges, different contribution of extended PWN emission (because of the limited telescope resolution), and some of them did not  isolate the nonthermal magnetospheric emission from a possible thermal component (generally seen below $\sim2$ keV).  Nonetheless, 
   \citet{2002A&A...387..993P} have already pointed out that the best-fit relation does not provide  a statistically acceptable fit to the data due to the very large scatter in \lpsrx for pulsars with similar $\dot{E}$ values and  noted that all \lpsrx points appear to lie below the  curve (upper bound) given by  \lpsrxc$\propto\edot^{1.48}$.  This conclusion  was strengthened by KP08 who collected 
    the results of \chan observations of $\sim40$ pulsars and their PWNe and found   \lpsrxc$\propto\edot^{1.3}$  and  \lpwnxc$\propto\edot^{1.6}$ (generally consistent with \citealt{2002A&A...387..993P}, who  did not separate the pulsar and PWN contributions).

After the launch of  the {\sl Fermi} $\gamma$-ray observatory, the number of $\gamma$-ray detected pulsars has  grown rapidly, and it has nearly matched the number of  X-ray detected pulsars after three years of LAT operation\footnote{See https://confluence.slac.stanford.edu/display/GLAMCOG/Public+List+of+LAT-Detected+Gamma-Ray+Pulsars}.  The achieved progress makes  it possible to carry out studies similar to those in X-rays. In particular,  \citet{2011ApJ...733...82M} have  studied both X-ray and $\gamma$-ray properties of 29 {\sl Fermi} pulsars with well-characterized X-ray spectra.  From analyzing the X-ray  properties  of these pulsars, \citet{2011ApJ...733...82M}  found  the best-fit correlation \lpsrx$\propto\edot^{1.04}$, albeit again with a large scatter, which made this fit formally  unacceptable.  The best-fit   correlation for the $\gamma$-ray luminosity, \lpsrg$\propto\edot^{0.88}$, also resulted in a poor   fit. However, in this case the poor quality of the  fit could be caused by an apparent  break  at $\edot_{\rm crit} \approx 3.7\times 10^{35}$ erg s$^{-1}$ rather than just by the  scatter.   Above  $\edot_{\rm crit} $, the best fit relation appears to be \lpsrg$\propto\edot^{0.2}$ while it is \lpsrg$\propto\edot^{1.43}$ below  $\edot_{\rm crit}$.   Marelli  et al.\ (2011) also considered the dependence of the distance-independent \lpsrg/\lpsrx ratio on $\edot$ and found that the ratio shows a strong scatter (up to 3 orders of magnitude) and  a very weak (or no) correlation with $\edot$ (see Figure 4 of \citealt{2011ApJ...733...82M}). The scatter in   \lpsrg/\lpsrx could simply be caused by the scatter in \lpsrx.

In this paper we present  analysis based on the largest reported sample of isolated, rotation-powered pulsars observed by the Advanded CCD Imaging Spectrometer (ACIS) aboard {\sl CXO}. The advantage of {\sl CXO}/ACIS is that  even within a short exposure it is possible to achieve  deep detection limits,  thanks to the very low ACIS background and sharp point spread function (PSF) of the telescope \citep{2003SPIE.4851...28G}.  Most of the pulsars  reported here were observed in the course of our guaranteed observation time (GTO) program  (PI G.~Garmire)  with $\approx$10\,ks ACIS exposures. The rest of the data were taken from the {\sl CXC} archive\footnote{\tt \href{http://cxc.harvard.edu/cda/}{http://cxc.harvard.edu/cda/}}. In the sample of  23 pulsars, 12 are detected by {\sl Fermi} LAT and listed in the 2FGL catalog \citep{2011arXiv1108.1435T} or reported elsewhere.  We also made use of  KP08  and \cite{2007ApJ...664.1072P} to include previously reported results.
  In Section \ref{obs} we describe how we measure the fluxes  and their upper limits. In  Section \ref{results} we provide the measured parameters for each pulsar, as well as derived quantities such as X-ray luminosities and  efficiencies. These results are discussed in Section \ref{disc}, where we combine our findings with the previous results in the X-ray range and compare these with the $\gamma$-ray properties. Finally, we present our conclusions in Section \ref{sum}.

\section{Observations and Analysis}\label{obs} 
 The fields of 23 pulsars were imaged with the ACIS I-array or S-array, with exposure times of typically 10\,ks, as part of our GTO program, between 2001 and 2011. 
 The data for each observation were processed 
 using the standard pipeline.
  We filtered the pipeline-produced event lists, keeping only photons with energies 0.5--8\,keV, and searched for X-ray sources in the vicinity of the known pulsar coordinates (see Figure 1). The coordinates, taken from the most recent ATNF catalog \cite{2005AJ....129.1993M}, 
  typically should have subarcsecond uncertainties (although there can be exceptions, see e.g.,   \citealt{2007ApJ...660.1413K}). The final positional uncertainty region in the image is the combination of the ATNF coordinate uncertainty with the typical {\sl Chandra} pointing error, 0\farcs6 at 90\% confidence. We then searched for significant X-ray detections within this area in each  ACIS image.

 We consider a target detected when, for $N$ detected counts, the probability  of finding  $\geq N$ events by chance within a chosen aperture  is less than 0.0001 ($\approx4\sigma$). For Poisson statistics,  this probability is given by  
\begin{equation}
	P(N,\lambda) = 1-e^{-\lambda}\sum_{i=0}^{N} \frac{\lambda^i}{i!},
\end{equation}
where $\lambda$ is the average number of background  counts in the source aperture. 
  In observations where a source was seen within the search area, we placed our measurement apertures at the centroid of the photon distribution; in the case of no detection, we placed our aperture at the centre of the search region. In each case the background was measured in much  larger regions free of sources, but located on the same chip.
  
 For our short exposures,   the numbers of detected photons 
   are typically  too small to perform a reliable spectral fitting. Therefore, we 
  adopted a more straightforward approach to 
 estimate the observed fluxes. 
   We used the CIAO task {\tt psextract} to 
   calculate the effective area, $A(E_i)$, at a given energy $E_i$ of the detected photon,  at the position of the source. 
     The observed flux and its uncertainty were then estimated following \cite{2009ApJ...691..458P}:
\begin{eqnarray}
f = T_{\rm exp}^{-1} \sum_i  E_i A(E_i)^{-1}, \label{flux}\\
\delta f = T_{\rm exp}^{-1}  \left[\sum_i E_i^2 A(E_i)^{-2}\right]^{1/2}, \label{dflux}
\end{eqnarray}
where $E_i$ is the energy of the $i$th photon and $T_{\rm exp}$ is the exposure time 
  (the sum of good times corrected for deadtime).
We measured the flux in both an $r=1$\arcsec\ aperture, appropriate for point sources (it contains $\approx93$\% of the point source flux  for photons with $E=1$\,keV) and in an $1'' \leq r \leq 3''$ annulus, to measure possible extended emission from a compact PWN. We  subtracted  5\% of the point source flux due to the wings of the 
PSF within the  $1'' \leq r \leq 3''$ annulus.

  In the case of  non-detection,  there is no an even crude measure of the spectrum to use in Equations (\ref{flux}) and (\ref{dflux}).   
 Therefore, to calculate an upper limit,  we calculate the number of counts corresponding to $P(N,\lambda)=0.1$ (i.e., establishing a 90\% confidence limit) for the measured background rate, and use webPIMMS\footnote{\tt http://heasarc.nasa.gov/Tools/w3pimms.html} to calculate the equivalent flux in the 0.5--8\,keV band  for a typical pulsar spectrum (a power-law with photon index $\Gamma$=1.5 and galactic absorption column appropriate to each source;  see Section \ref{results}).
 
\section{Results}\label{results}
The immediate  results of our analysis are summarize in Table \ref{yes}  (detections) and Table \ref{no}  (non-detections). 
 Although the detection significance is high for every source in Table \ref{yes}, 
  the flux measurements 
  can be rather uncertain. For instance, it may be that several low- or mid-energy photons establish the detection, but the measured flux is dominated by a single high-energy photon (where the detector is much less sensitive). In such cases  the flux uncertainty will be of the order of the measured flux.

The only dubious case is B1822$-$14, which has an excess of counts over the background in both data-sets but no apparent point-like source. The excess counts could, in principle, be due to  a PWN 
 with an approximate flux of 3$\times10^{-15}$\,erg\,s$^{-1}$cm$^{-2}$, close to the limit we derive. Although two significant X-ray sources happen to fall on the same S3 chip, their positions are inconsistent with that of B1822$-$14, and both sources have 2MASS counterparts. Also, one additional ATNF pulsar is in the field (J1837$-$0604), which does not have an X-ray counterpart. It is included as a non-detection in Table \ref{no}.

Note that  there are faint X-ray sources in the vicinity of PSRs J1105$-$6107 and J1730$-$3350, 
 but in both cases they are  too far away to be acceptable pulsar counterparts.

Table \ref{psr} lists the dispersion measure (DM), spin-down luminosity $\dot{E}$, and distance $d$, of the pulsars, taken from the ATNF catalog\footnote{\tt http://www.atnf.csiro.au/research/pulsar/psrcat/expert.html}. Using these values, we can estimate the absorption column, $N_{\rm H}=3.1\times10^{19}$DM\,cm$^{-2}$ 
   (assuming an average  10\% degree of ionization along the line of sight), unabsorbed flux, luminosity, and  
efficiency.

For three pulsars, J1958+2846, J1413$-$6205 and J1023$-$5746, DM is not known. 
 In these cases we assumed a distance based on the position of the most prominent spiral arms in the direction to the pulsar, and corresponding extinction values. The calculated luminosity and efficiency values must therefore be taken with particular caution.

\begin{figure*}
\begin{center}
\includegraphics[width=17cm,height=16cm]{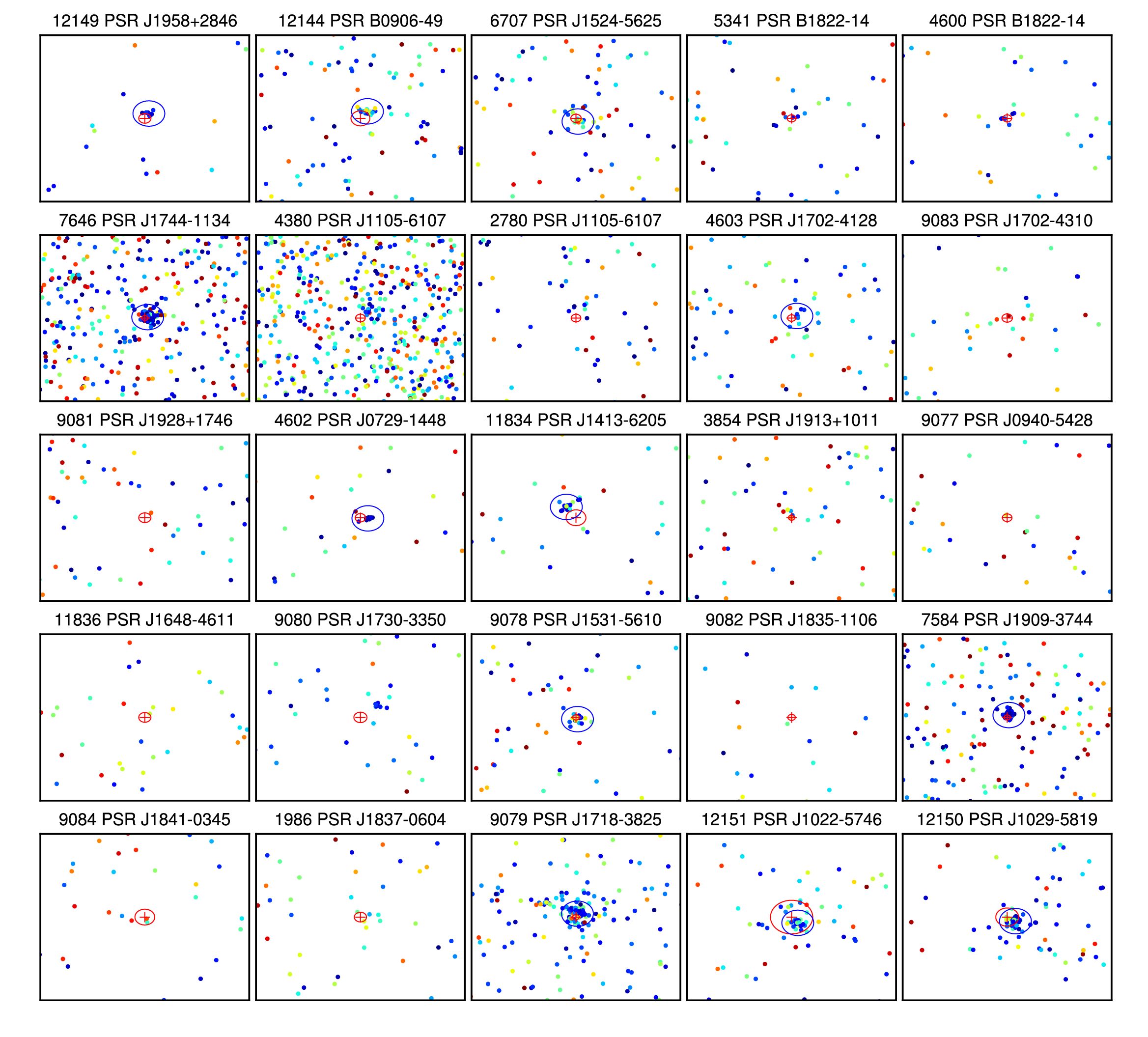}
\caption{Scatter plots showing the distributions of photons and their energies  for  the pulsar fields. Each panel is 40\arcsec$\times$40\arcsec. North is up, East is to the left. The positional uncertainties are shown by red circles (centred on the ATNF pulsar positions shown by the red crosses); a 3\arcsec\ aperture is shown for detected sources. Each photon is color coded according to its energy. }\label{scatter}
\end{center}
\end{figure*}

\section{Discussion}\label{disc}

\begin{figure*}
\begin{center}
\includegraphics[width=10.0cm,angle=90]{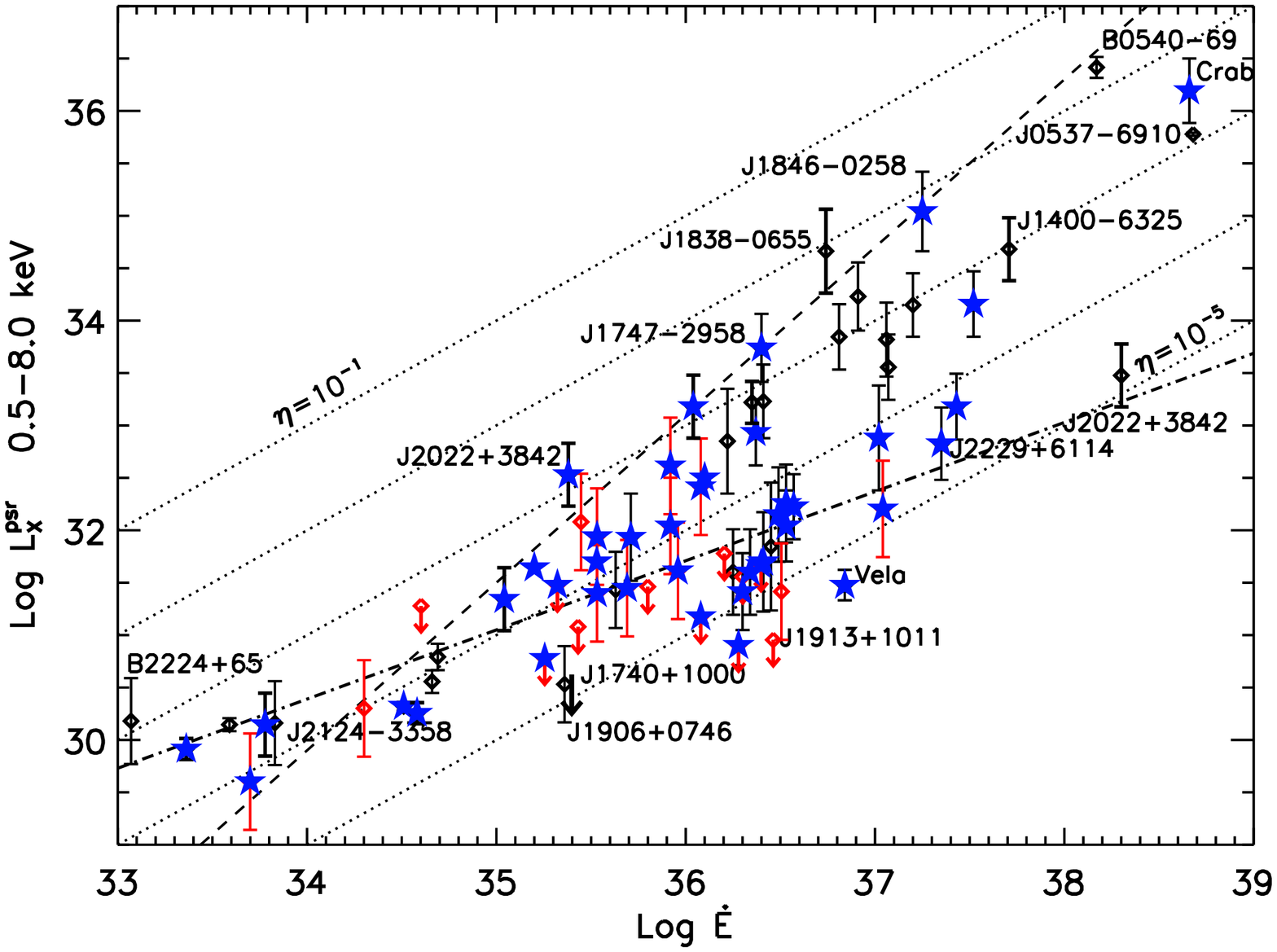}
\includegraphics[width=10.0cm,angle=90]{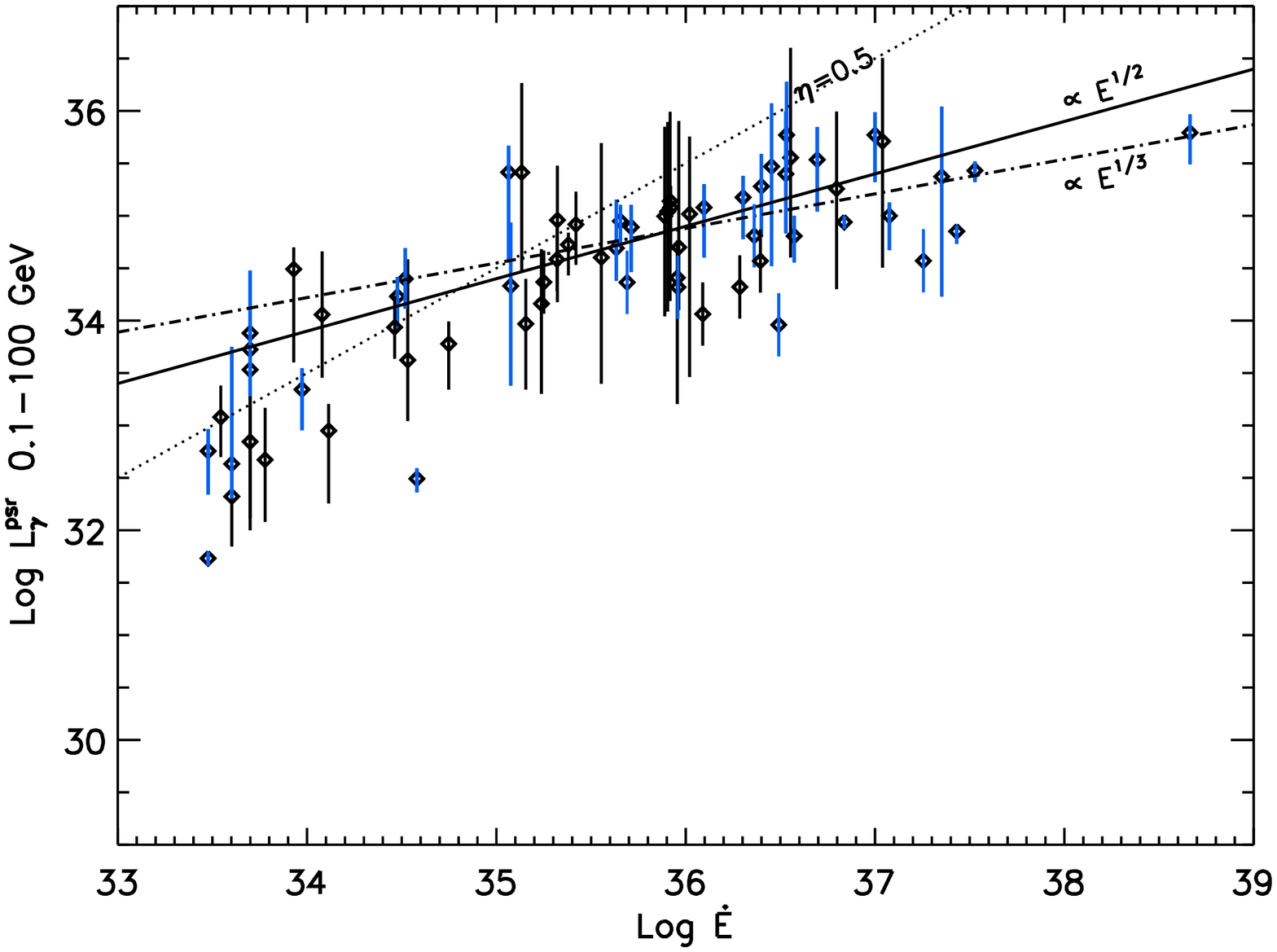}
\caption{\footnotesize  The top panel shows non-thermal X-ray luminosity vs.\ spin-down power $\dot{E}$.  The dashed and dash-dotted lines corresponds to $L_X\propto E^{1.6}$ and $L_X\propto E^{2/3}$, respectively (see text for discussion). The constant efficiency ($\eta=L/\dot{E}$) lines are shown by the dotted lines. The downward  arrows show $90$\% confidence  upper limits. The blue stars mark $\gamma$-ray pulsars.  The red errorbars and limits are from this paper, the rest are taken from KP08. The bottom panel shows $\gamma$-ray  luminosity (in 0.1--100 GeV) vs. pulsar's $\dot{E}$.  
 X-ray detected pulsars are shown in blue. 
  }\label{spec:xray}
\end{center}
\end{figure*}

By adding the flux measurements or upper  limits for 23  pulsars observed with  {\sl Chandra} ACIS we have significantly expanded the  sample of pulsars  
  analyzed by KP08.  In Figure \ref{spec:xray} (top panel) we plot the pulsar luminosity, \lpsrx, versus spin-down power $\edot$.  One can see that,  in general, \lpsrx increases with $\edot$, in  agreement with the previously noticed trends  (e.g.,  \citealt{1988ApJ...332..199S,1997A&A...326..682B,2002A&A...387..993P,2004ApJ...617..480C,2008ApJ...682.1166L}; KP08).  However, the correlation is rather weak, and, because of  the very large dispersion, it cannot be meaningfully described by a simple dependence \lpsrx$(\edot)$.   The large scatter is also manifest in   X-ray efficiencies,   
 \etapsrx=\lpsrx/$\edot$, which range from $\sim 10^{-5.5}$ to $\sim 10^{-1.8}$ in Figure \ref{spec:xray} (top). 
For instance, such well-known pulsars as the Crab and B0540-69 are very efficient, while  the very young, high-$\edot$ pulsar  $J2022+3842$, recently discovered by  \cite{2011ApJ...739...39A}, has 
   \etapsrx$=10^{-5}$ (for a plausible  $d=8$ kpc). Similarly large variations in \etapsrx are seen at lower $\edot$ down to $\edot\sim10^{36}$ erg s$^{-1}$.

 An obvious cause of the scatter could be incorrectly determined distances for some of the pulsars. However, for majority of them (including J2022+3842), the distances cannot be wrong by more than a factor of a few, too little to explain the scatter. One can allude to the beaming of the magnetospheric radiation as another possible factor 
 contributing to the spread in \etapsrx. However, the X-ray efficiencies of PWNe accompanying many of these pulsars show a similarly large scatter  (see Figure 1 in  KP10), although the PWN emission is not expected to be substantially beamed.   
 In several cases neither pulsar nor PWN 
were detected, including the most X-ray underluminous PSR
J1913+1011, for which \etapsrx$+$\etapwnx$ < 5.8\times 10^{-6}$. (Note, that the limit also includes any thermal pulsar emission and compact PWN contribution, which means that the actual limit on the nonthermal magnetospheric emission must  be even lower.) 
 Therefore,  the lack of tight correlation between the \lpsrx and $\edot$ can hardly be   explained just by  the beaming and poorly known distances.

 Despite the huge scatter, the  {\em maximum upper bound}  on  \lpsrx at given $\dot{E}$ appears to be well defined. For $\dot{E}\gtrsim 10^{35}$ erg s$^{-1}$
  it  approximately follows \lpsrxc$\propto E^{1.6}$ (dashed line in Figure \ref{spec:xray}, top panel); however,  for $\dot{E}\lesssim 10^{34}$  erg s$^{-1}$  the  dependence on $\dot{E}$ seems to  flatten to $\propto \dot{E}^{2/3}$ or even flatter \citep{Posselt2012}.
 
Both the extreme scatter and the existence of the upper bound  suggest that  additional parameters must enter in the \lpsrx$(\edot,...)$ dependences.  One possibility  is that  there may be two qualitatively different emission regimes  which  correspond to two distinct  \lpsrx$(\edot)$ (e.g., those shown by the dashed and dash-dotted lines in  Figure \ref{spec:xray}, top panel). Given all the uncertainties mentioned above, the current data could be consistent with such a dichotomy although 
  other alternatives, such as a continuous  dependence  of \etapsrx on some parameter, are also possible.  This parameter, however,  is unlikely to be just the angle between the magnetic dipole and pulsar spin axis because the orthogonal rotator B0906--49 \citep{2008MNRAS.390...87K} has an unremarkable X-ray efficiency
  compared to other pulsars with similar $\edot$. 
    One can also speculate that for low-$\dot{E}$ pulsars 
  the PWN becomes so compact that it is cannot be resolved even with {\sl CXO}. 
   A larger sample of pulsars with well known distances and good quality spectra is required to discriminate between the various alternatives. Also, the measurements of  pulsed non-thermal emission  can be used to constrain the very compact PWN contribution.

 \begin{figure*}
\includegraphics[width=13cm,angle=90]{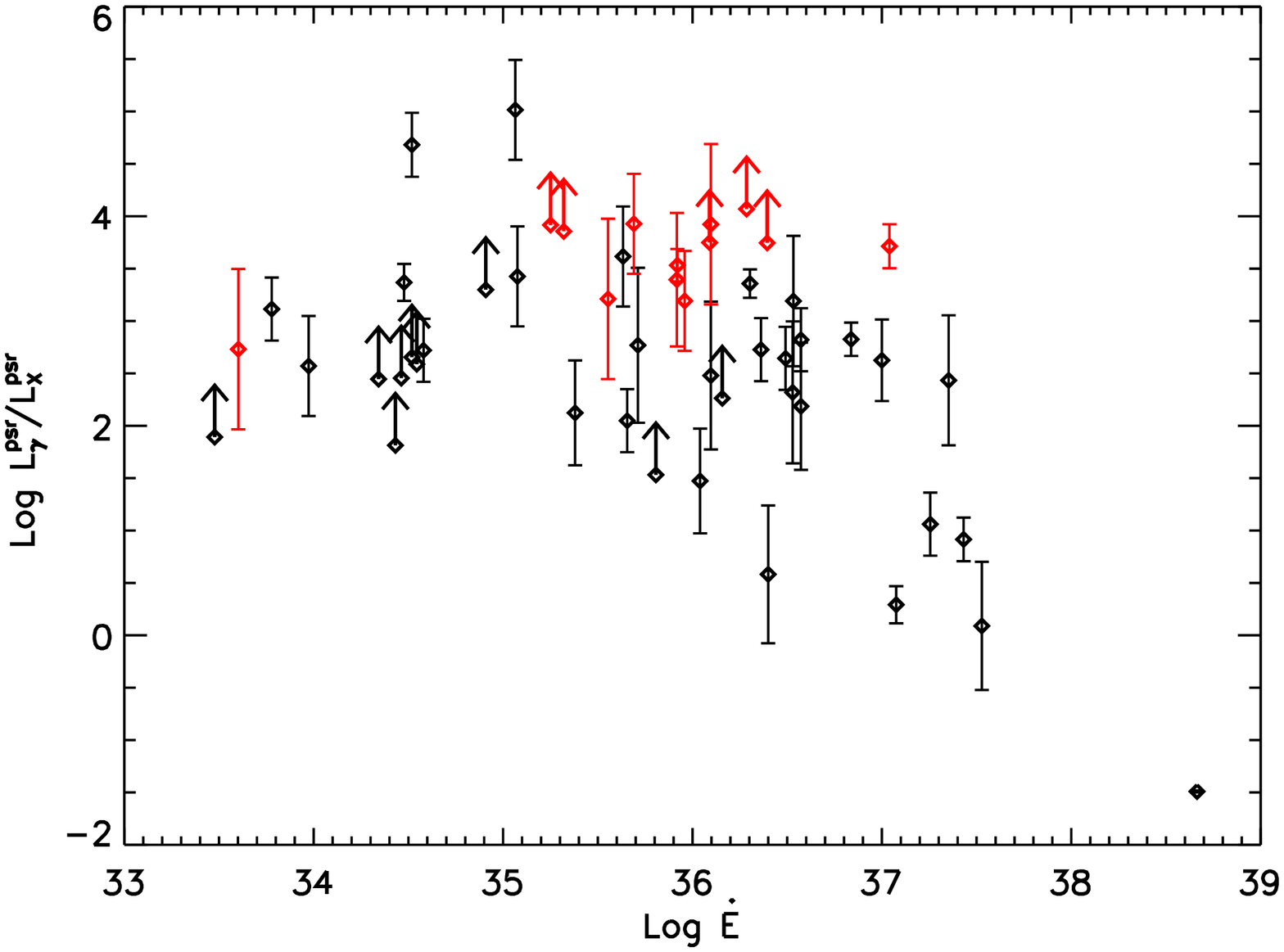}
\caption{X-ray to $\gamma$-ray flux ratios for {\sl Fermi/}LAT-detected pulsars that have been observed with {\sl CXO}/ACIS. The ratio is independent of  distances, which are poorly known for many pulsars. The pulsars analyzed in this paper are shown in red.}\label{X_G}
\end{figure*}

 It is interesting to compare the  \lpsrx vs.\ $\edot$  correlation with that seen in $\gamma$-rays. \cite{2010ApJS..187..460A} have presented the first analysis of the properties of 46  pulsars detected by {\sl Fermi} LAT. Shortly afterwards, more {\sl Fermi} pulsars were discovered \citep{2010ApJ...725..571S, 2012ApJ...744..105P}. 
 We have calculated the $ >0.1$ GeV luminosities  of the 54 $\gamma$-ray pulsars  using the published\footnote{If no published flux values were found, we took them from the 2FGL catalog (\citealt{2011arXiv1108.1435T}). } pulsar fluxes and the best published estimates of the distances. 
  Figure \ref{spec:xray} (bottom panel) shows the  correlation between the $\gamma$-ray luminosity, \lpsrg, and $\dot{E}$ (cf.\ Figure 2 in \citealt{2011ApJ...733...82M}).     The correlation between \lpsrg and $\edot$ appears to be tighter  than that between  \lpsrx and\ $\edot$ (despite the presumably stronger beaming in $\gamma$-rays as follows from higher pulsed fractions; \citealt{2010ApJS..187..460A}), 
 and it also  differs in shape from the  \lpsrx vs.\ $\dot{E}$  correlation\footnote{Note that the vertical axis range is the same in both panels of Figure \ref{spec:xray}.}  (cf.\ top and bottom panels in Figure \ref{spec:xray}).
 For the energetic pulsars ($\dot{E}\gtrsim10^{35}$ erg s$^{-1}$)  the  \lpsrg vs.\ $\dot{E}$  correlation appears to be consistent with the expected $\dot{E}^{1/2}$ law (e.g., \citealt{1981ApJ...245..267H}), or with an even flatter one (see  \lpsrg$\propto \dot{E}^{1/3}$ line in Figure 3, bottom).  However, at lower $\dot{E}$ the observed correlation is more consistent with an  \lpsrg$\propto\dot{E}$ scaling, implying a break around $\edot\sim10^{35}$ erg s$^{-1}$ in the \lpsrg$(\dot{E})$ dependence.   The break  hints at a qualitative change  either in the emission mechanism or in the spectral energy distribution (SED) 
   of the primary particles (see below). 
     Such a break was  
     expected to occur  at somewhat lower $\dot{E}\simeq10^{33}$ erg s$^{-1}$ in  the polar cap  model  \citep{2002ApJ...576..366H}. However,  recent simulations  based on the slot gap model seem to predict a break in  the \lpsrg--$\dot{E}$ relationship at $\dot{E}\simeq10^{35}$ erg s$^{-1}$ (see Figure 1 in \citealt{2011AIPC.1357..249P}).
        It is  more difficult to 
 determine how the break in the primary particle SED will affect the properties of the secondary electrons and their synchrotron emission (see below), but  some impact is likely, and it could be seen in the \lpsrx--$\dot{E}$ relationship (Figure \ref{spec:xray}, top panel) and in the multiwavelength spectra.
 Indeed, there is an intriguing coincidence between the values of $\dot{E}$
  at which the \lpsrxc--$\dot{E}$ and \lpsrg--$\dot{E}$ appear to exhibit a break, although  the slopes change in the opposite ways.
  Also, the  \lpsrg\ vs.\ $\dot{E}$ correlation is significantly stronger (i.e.,  the scatter is weaker) than the  \lpsrx vs.\ $\edot$ correlation.
 
  The comparison of the  top and bottom panels in Figure \ref{spec:xray} makes it obvious that while the X-ray and $\gamma$-ray efficiencies can be similar for some very young pulsars,    older pulsars are generally more efficient $\gamma$-ray emitters. This can also be seen in Figure \ref{X_G},  where we plot the  distance-independent X-ray to $\gamma$-ray luminosity ratio for gamma-ray pulsars observed in X-rays. Although the scatter is large (mainly due to that in \lpsrx),
 the ratio increases with decreasing $\edot$ down to $\edot\simeq10^{35}$ erg s$^{-1}$, at which point the trend appears to reverse. These results suggest $\edot$-dependent changes in the shapes of the multiwavelengths (MW) spectra of pulsars.  Perhaps, it could be a change in the slope of a broadband power-law (PL) spectrum (if one attempts 
 to describe most of MW emission as a curved or broken PL) or more complex changes. 
   To better understand  the implications of our findings for the magnetospheric models, a larger number of $\gamma$-ray pulsars with $\dot{E}=10^{33}-10^{35}$ erg s$^{-1}$  should be observed in  X-rays with long enough exposures to either detect their X-ray emission or  set  restrictive upper limits.

It is generally believed that the pulsar GeV emission  is produced by the curvature radiation of primary electrons pulled from the   neutron star (NS) surface and accelerated by the electric field component parallel to the magnetic field.  The curvature photons initiate pair cascades  leading to the production of secondary  or higher-generation electrons that emit synchrotron radiation at lower frequencies  (optical to X-rays; e.g., \citealt{1998ApJ...493L..35C,2008AIPC..968..104H,2009ASSL..357..521H}, and reference therein).   In the high-altitude slot-gap and outer-gap models the  primary particles keep accelerating up to 10--100 NS radii,  gain  momentum  transverse  to the magnetic field via resonant cyclotron absorption \citep{1998A&A...337..433L}, and emit significant synchrotron radiation up  to MeV (and possibly even GeV) energies  in young pulsars (e.g.,  \citealt{2011AIPC.1379...74B}).  Within this framework, it is still possible to have a MW (from optical to GeV) spectrum whose shape would resemble a single broadband PL with a cutoff at the highest energies \citep{2008MNRAS.386..748T}, which seems to be in qualitative agreement with MW spectra of some pulsars \citep{2011ApJ...743...38D}.    However, additional processes, such as the modification of the spectrum by inverse Compton scattering \citep{2008AIPC..968..104H} or by synchrotron self-Compton process \citep{2002ApJ...569..872Z}, can play important role under certain conditions.  Our findings imply that the relative contributions  of these processes may vary, depending on the geometry of the magnetosphere 
  or   the $\dot{E}$ magnitude.

\section{Summary}\label{sum}

By analyzing the population of rotation-powered pulsars detected by {\sl CXO},  we found that  the \lpsrx--$\edot$ relationship cannot be meaningfully described as a simple \lpsrx$(\edot)$ dependence.  There is some degree of correlation  between \lpsrx and $\edot$, but the extreme scatter (by $>4$ orders of magnitude) in the X-ray radiative efficiencies is present for pulsars with $\edot\gtrsim10^{36}$ erg s$^{-1}$. Although existing data hint that the scatter may decrease with decreasing $\edot$, perhaps becoming substantially smaller at  $\edot\lesssim10^{35}$ erg s$^{-1}$, this could simply be the result of small-number statistics  (few pulsars with low $\edot$ have been detected)  and of the limited sensitivity of  existing observations. 
  The reasons for the scatter are unclear.    The deepest X-ray limits reported in this paper strongly support  
  the idea that  \lpsrx depends not only on $\edot$ but also strongly depends on other parameters.
  At the same time, it seems that the upper bound  on  the \lpsrx--$\edot$ relationship is fairly well defined (\lpsrxc$\propto E^{1.6}$), corresponding to such values of the ``hidden'' parameters that deliver maximum radiative efficiency at a given $\edot$.
 
 We thank Brian Newman who participated in the initial stages of this work. The comparison with a sample  of $\gamma$-ray pulsars detected by {\sl Fermi} LAT  shows that the correlation  between   \lpsrg  and  $\edot$ is much tighter, but again it  can hardly be described by a simple power-law  \lpsrg$(\edot)$ dependence. The break  between $\edot\sim10^{34}$ and $10^{35}$ erg s$^{-1}$  is suggested by the existing data.   Intriguingly, a break  in  \lpsrxc--$\edot$ is at  similar $\edot$ values. 

\medskip\noindent

The work on this project was partly supported through the NASA
grants NNX06AG36G, NNX09AC81G and NNX09AC84G, and National Science Foundation
grants No.\ AST0908733 and AST0908611. The work by G. G. P. was partly supported by the Ministry
of Education and Science of the Russian Federation (contract 11.634.31.0001).

\bibliographystyle{apj}  
\bibliography{database}

\clearpage

\begin{deluxetable}{lccccccccccc}
\tablecaption{Pulsar detections.\label{yes}}
\tablewidth{0in}
\tabletypesize{\footnotesize}
\tablehead{
\colhead{Object} & \colhead{ObsID} & \colhead{$T_{\rm exp}$ } & \colhead{Chip} & \colhead{R.A.} & \colhead{decl.} & \colhead{$N_{\rm ph} (1\arcsec)$} & \colhead{$\tilde{E}$} & \colhead{$N_{\rm ph} (3\arcsec)$} & \colhead{$N^{\rm bg}_{\rm ph}(1\arcsec)$} & \colhead{$F_{\rm X, obs}^{\rm psr}$} & \colhead{$F_{\rm X, obs}^{\rm pwn}$}\\
 & & ks & & deg& deg&& keV &&&
}
\startdata 
PSR J1958+2846   & 12149 & 9.9 & 2 & 299.66672 &   +28.76531 & 10 &1.3& 12 & 0.1& 13.6$\pm$4.4 & 4.0$\pm$3.0\\
PSR B0906$-$49   & 12144 & 34.6& 3 & 137.14747 & $-$49.21801 & 10 &2.9& 17 & 0.4& 4.2$\pm$1.6 & 3.4$\pm$1.5\\
PSR J1524$-$5625 & 6707  & 13.7& 7 & 231.20756 & $-$56.42337 & 9  &3.9& 16 & 0.4& 14.4$\pm$6.1 & 24$\pm$19\\
PSR J1744$-$1134 & 7646  & 63.3& 7 & 266.12236 & $-$11.58179 &273 &1.1& 312& 1.6& 16.8$\pm$1.1& 13.8$\pm$6.6\\
PSR J1702$-$4128 & 4603  & 10.4& 7 & 255.71846 & $-$41.47992 &  8 &2.2& 13 & 0.3 & 22.4$\pm$17.1&11.9$\pm$9.1\\
PSR J0729$-$1448 & 4602  & 4.1 & 7 & 112.31814 & $-$14.81026 & 13 &0.9& 15 & 0.2& 10.2$\pm$2.9 & 64$\pm$63\\
PSR~J1413$-$6205 & 11834 & 9.9 & 0 & 213.37566 & $-$62.09318 & 8  &1.5& 14 & 0.1 & 8.9$\pm$3.7 & 7.2$\pm$3.4\\
PSR J1531$-$5610 & 9078  & 9.9 & 3 & 232.86615 & $-$56.18206 & 11 &1.9& 17 & 0.1& 34.0$\pm$22.8 & 8.2$\pm$3.8\\
PSR J1909$-$3744 & 7584  & 29.7& 7 & 287.44754 & $-$37.73717 & 64 &1.0& 71 & 0.8& 7.9$\pm$1.0& 0.9$\pm$0.3\\
PSR J1718$-$3825 & 9079  & 9.9 & 3 & 259.55642 & $-$38.42146 & 36 &1.6& 79 & 0.2& 43.9$\pm$11.1 & 72$\pm$24\\
PSR J1022$-$5746 & 12151 & 9.9 & 3 & 155.76147 & $-$57.76843 & 24 &2.9& 35 & 0.1& 31.4$\pm$7.2& 14.7$\pm$4.9\\
PSR J1028$-$5819 & 12150 & 9.9 & 2 & 157.11584 & $-$58.31837 & 45 &1.2& 61 & 0.1 & 40.9$\pm$7.4& 19.3$\pm$5.5\\
\enddata
\tablecomments{ Chips 0--3 are ACIS-I chips, and chip 7 is ASIC-S3. $N_{\rm ph}(1'')$ is the number of counts within the $r=1''$ circle centered on the X-ray source,  and $\tilde{E}$ is the median energy of these photons.  Positional uncertainties of the X-ray sources are 
   dominated by the  {\sl Chandra} pointing error (0\farcs3 at 68\% confidence).  $N_{\rm ph}(3'')$ is the number of counts in the $r=3''$ circle.  $N^{\rm bg}_{\rm ph}(1\arcsec) $  is the expected mean number of background counts scaled to the $r=1''$ aperture.    $F_{\rm obs}^{\rm psr}$ and $F_{\rm obs}^{\rm pwn}$ are the  observed (absorbed) pulsar and PWN fluxes in  the 0.5--8\,keV band, in units 10$^{-15}$\,erg\,s$^{-1}$cm$^{-2}$.
  $F_{\rm obs}^{\rm PWN}$ is the  observed 
   flux in 
   the $1'' \leq r \leq 3''$ annulus, after subtracting the flux due to the point source. }
\end{deluxetable}

\begin{deluxetable}{lcccccccccc}
\tablecaption{Pulsar non-detections.\label{no}}
\tablewidth{0in}
\tabletypesize{\footnotesize}
\tablehead{
\colhead{Object} & \colhead{ObsID} & \colhead{$T_{\rm exp}$ (ks)} & \colhead{Chip} & \colhead{R.A.} & \colhead{decl.} &   \colhead{$N_{\rm ph} (3\arcsec)$} & \colhead{$N^{\rm bg}_{\rm ph}(3\arcsec)$}  & \colhead{$F_{\rm X, u.l.}$} 
}
\startdata 
PSR B1822$-$14   & 5341 & 18.0 & 0 & 276.26220 & $-$14.78127 & 7 & 1.9 & 3.7\\
 PSR B1822$-$14                & 4600 & 11.0 & 0 &   276.26220        &     $-$14.78127        & 6 & 1.1 & 3.0\\
PSR J1105$-$6107 & 4380\tablenotemark{a} & 11.7 & 7 & 166.35904 & $-$61.13094 & 15& 24 & ...\\
  PSR J1105$-$6107               & 2780 & 11.6 & 7 &   166.35904          &   $-$61.13094          & 5 & 2.5 & 5.6\\
PSR J1702$-$4310 & 9083 & 9.6  & 3 & 255.61225 & $-$43.17778 & 3 & 1.3 & 5.3\\
PSR J1928+1746   & 9081 & 9.9  & 3 & 292.17700 & +17.77417    & 1 & 2.1 & 5.8\\
PSR J1913+1011   & 3854 & 19.5 & 7 & 288.33475 & +10.18971    & 5 & 2.9 & 2.9\\
PSR J0940$-$5428 & 9077 & 10.0 & 3 & 145.24258 & $-$54.47794 & 1 & 1.1 & 2.8\\
PSR J1648$-$4611   & 11836& 10.0 & 2 & 252.09175 & $-$46.18778 & 2 & 1.3 & 5.2\\
PSR B1727$-$33   & 9080 & 9.9  & 3 & 262.63566 & $-$33.84428 & 1 & 1.7 & 4.8\\
PSR J1835$-$1106 & 9082 & 10.0 & 2 & 278.82620 & $-$11.10419 & 1 & 1.2 & 4.2\\
PSR J1841$-$0345 & 9084 & 10.0 & 3 & 280.41117 & $-$3.81183  & 3 & 1.3 & 4.4\\
PSR~J1837$-$0604 & 1986 & 9.0  & 6 & 279.43146 & $-$6.08028 & 4 & 1.5 & 5.1\\
\enddata
\tablecomments{The coordinates correspond to the  radio pulsar positions from ATNF catalog \cite{2005AJ....129.1993M}. Chips 0--3 are ACIS-I, and chip 7 is ASIC-S3. $N_{\rm ph}(3'')$ is the number of counts within the $r=3''$ circle centered on the radio pulsar position.  $N^{\rm bg}_{\rm ph}(3\arcsec) $  is the expected mean number of background counts scaled to $r=3''$ circular aperture. $F_{\rm X, u.l.}$ is the upper limit on the observed (absorbed)  flux in the 0.5--8\,keV range corresponding to 90\% confidence limit (see Section 2), in units of 10$^{-15}$\,erg\,s$^{-1}$cm$^{-2}$.  } 
\tablenotetext{0}{$^{\rm a}$ This observation does not provide a useful limit because it was affected by a high background.}
\end{deluxetable}

\begin{deluxetable}{lcccccccccccc}
\tablecaption{Pulsar parameters.\label{psr}  }
\tablewidth{0in}
\tabletypesize{\footnotesize}
\tablehead{
\colhead{Object} &\colhead{$l$} & \colhead{$b$} & \colhead{DM}&   \colhead{P}&   \colhead{Age}& \colhead{$\dot{E}/10^{35}$}  & \colhead{$N_{\rm H}/10^{21}$} & \colhead{$d$} & \colhead{$F_{\rm X}/10^{-15}$\tablenotemark{a}} & \colhead{$L_{\rm X}/10^{31}$\tablenotemark{b}} & \colhead{$\eta_{\rm X}$\tablenotemark{c}}& \colhead{LAT\tablenotemark{d}}\\
 &deg&deg& \colhead{pc\,cm$^{-3}$} & ms & kyrs & \colhead{erg\,s$^{-1}$} & \colhead{cm$^{-2}$} & \colhead{kpc} & \colhead{c.g.s.} & \colhead{erg\,s$^{-1}$} 
}
\startdata 
PSR J1958+2846   & 65.88&$-$0.35&  ...      & 290& 22 & 3.4 & 5\tablenotemark{e}& 2\tablenotemark{e} & 52 & 2.5 & $7.4\times10^{-5}$ & Y\\
PSR B0906$-$49   &270.27&$-$1.02& 180& 107 & 112 &4.9 & 5.6&6.66 & 5.2 & 2.8 & $5.7\times10^{-5}$& Y\\
PSR J1524$-$5625 &323.00&   +0.35& 153 & 78 &32 & 32 & 4.7&3.84 & 15 & 2.6  & 8.1$\times10^{-6}$&  N\\
PSR J1744$-$1134 & 14.79&   +9.18&  3.14&  4 & $7\times10^{6}$ & 0.05 & 0.1&0.42& 17& 0.04 & 8$\times10^{-5}$ & Y\\
PSR J1702$-$4128 &344.74&   +0.12& 367 & 182 & 55 & 3.4 & 11.3&5.18& 27& 8.7 & 2.6$\times10^{-5}$ & N \\
PSR J0729$-$1448 &230.39&   +1.42&  92 & 252 & 35 & 2.8 & 2.8&4.37 & 52& 12 & 4.2$\times10^{-4}$ & N\\
PSR J1413$-$6205 &312.37&$-$0.74&  ...  & 110 & 63   & 8.3 & 10\tablenotemark{e}& 4\tablenotemark{e} & 215 & 41.1 &5.0$\times10^{-4}$&Y\\
PSR J1531$-$5610 &323.90&   +0.03& 111 & 84 & 97 & 9.1 & 3.4&3.10 & 36 & 4.1 & 4.5$\times10^{-5}$ & Y\\
PSR J1909$-$3744 &359.73&$-$19.60&  10.4& 3 & $3\times10^{6}$ & 0.2 & 0.3&1.27& 9.0& 0.2 & $1\times10^{-4}$ & N\\
PSR J1718$-$3825 &348.95&$-$0.43& 247 & 75 & 89 & 12 & 7.6&4.24 & 120 & 26 & $2.2\times10^{-4}$ & Y\\
PSR J1023$-$5746 &284.17&$-$0.41&   ...    & 111 & 46 & 110 & 15\tablenotemark{e} & 4.5\tablenotemark{e} & 64& 16 &1.5$\times10^{-5}$&Y\\
PSR J1028$-$5819 &285.06&$-$0.50&  96& 91 & 90 & 8.3 & 3.0&2.76 & 120& 11 & $1.2\times10^{-4}$ & Y\\
\hline
PSR B1822$-$14   & 16.81&$-$1.00& 357      & 279 & 195  & 0.4 & 11.0&5.45 &$<$5.5 & $<$1.9 & $<$4.9$\times10^{-4}$ & N\\
PSR J1105$-$6107 &290.49&$-$0.85& 271&  63 & 63 & 25 & 8.4&7.07 & $<$7.9 & $<$4.7 & $<$1.9$\times10^{-5}$ & Y\\
PSR J1702$-$4310 &343.35&$-$0.85& 377   & 240 & 17 & 6.3 & 11.6&5.44 &$<$8.0& $<$2.9 & $<$4.5$\times10^{-5}$ & N\\
PSR J1928+1746   & 52.93&   +0.11& 177 & 69 & 82 &16 & 5.5&8.13 & $<$7.7 & $<$6 & $<$3.8$\times10^{-5}$ & N\\
PSR J1913+1011   & 44.48&$-$0.17& 179 & 36& 169 & 29 & 5.5&4.48 & $<$3.9  & $<$0.9 & $<$3.3$\times10^{-6}$ & N\\
PSR J0940$-$5428 &277.51&$-$1.29& 134 & 87 & 42 & 19 & 4.2&4.27 & $<$3.5 & $<$0.8 & $<$4.0$\times10^{-6}$ & Y\\
PSR J1648$-$4611 &339.44&$-$0.79& 393 & 165 & 110 & 2.1 & 12.1&5.71&$<$7.8 & $<$3.0 & $<$1.5$\times10^{-4}$ & Y\\
PSR B1727$-$33   &354.13&   +0.09& 259   & 139 & 26 & 12 & 8.0&4.26 & $<$6.6 & $<$1.5 & $<$1.2$\times10^{-5}$ & Y\\
PSR J1835$-$1106 & 21.22&$-$1.51& 133  & 166 & 128 & 1.8 & 4.1&3.08& $<$5.2 & $<$0.6 & $<$3.4$\times10^{-5}$ & N \\
PSR J1841$-$0345 & 28.42&   +0.44& 194  &204 & 56 & 2.7 & 6.0&4.15& $<$5.9 & $<$1.2 & $<$4.5$\times10^{-5}$ & N\\
PSR J1837$-$0604 & 25.96&   +0.26& 462  & 96  & 34   & 20 & 14.3&6.19& $<$8.0 & $<$3.6 & $<$1.9$\times10^{-5}$ & N\\
\enddata
\tablenotetext{0}{$^{\rm a}$ Unabsorbed flux in 0.5--8 keV for the point source (measured from $r=1\arcsec$ circular aperture), corrected for the finite aperture size. For non-detections, the fluxes correspond to  90\% confidence  limits.}
\tablenotetext{0}{$^{\rm b}$ X-ray luminosity in 0.5--8 keV.}
\tablenotetext{0}{$^{\rm c}$ X-ray radiative efficiency $\eta_{X}=L_{X}/\edot$.}
\tablenotetext{0}{$^{\rm d}$ {\sl Fermi} LAT detection (Yes or No).}
\tablenotetext{0}{$^{\rm e}$ Crude estimates for the distance and $N_{\rm H}$, to be regarded with caution. }
\end{deluxetable}

\end{document}